# Spatially evolving vortex-gas turbulent free shear layers:
# Part 1. Effect of velocity ratio, and upstream and downstream conditions on spread-rate.

Saikishan Suryanarayanan and Roddam Narasimha

*Engineering Mechanics Unit, Jawaharlal Nehru Centre for Advanced Scientific Research, Jakkur, Bangalore – 560064, India*

The relevance of the vortex-gas model to the large scale dynamics of temporally evolving turbulent free shear layers in an inviscid incompressible fluid has recently been established by extensive numerical simulations (Suryanarayanan et al, *Phys. Rev. E* 89, 013009, 2014). Here, the effects of the velocity ratio across a spatially evolving 2D free shear layer are investigated by vortex-gas simulations, using a computational model based on Basu et al (1992, 1995), but with a crucial improvement that ensures conservation of global circulation. These are carried out for a range of values of the velocity ratio parameter $\lambda = (U_1 - U_2)/(U_1 + U_2)$, where $U_1$ and $U_2$ ($< U_1$) are respective velocities across the layer. The simulations show that the conditions imposed at the beginning of the free shear layer and at the exit to the domain can affect the flow evolution in their respective neighborhoods, the latter being particularly strong as $\lambda \to 1$. In between the two neighborhoods is a regime of universal self-preserving growth rate given by a universal function of $\lambda$. The computed growth rates are located within the scatter of experimental data on plane mixing layers, and in close agreement with recent high Reynolds number experiments and 3D LES studies, past the mixing transition, and support the view that free shear layer growth can be largely explained by the 2D vortex dynamics of the quasi-two-dimensional large scale structures.

## I. INTRODUCTION

The incompressible turbulent mixing layer or free shear layer is a canonical flow of great importance, both because of its role as perhaps the simplest conceivable turbulent shear flow and because of its frequent occurrence in numerous applications. A large body of experimental data is available on the behavior and character of such flows (e.g. Liepmann & Laufer 1947, Wygnanski et al, 1970, Winant & Browand 1974, Brown & Roshko 1974; most recently D'Ovidio & Coats 2013; see Brown & Roshko 2012 for a review). There have been several computational studies of the free shear layer, both 2D vortex simulations (e.g. temporal : Aref & Siggia, 1980; spatial: Ashurst 1979, Inoue 1985, Inoue & Leonard 1987, Ghoniem & Ng 1987, Basu et al, 1992, 1995) as well as more recent direct numerical simulations (DNS) of 3D Navier-Stokes equations (e.g. temporal: Rogers & Moser, 1994; spatial: Wang et al, 2008). However, there are still several important questions that remain unresolved. For example :

- How does the growth rate depend on initial and downstream conditions? Is there a universal growth rate regime independent of initial and boundary conditions for each velocity ratio? If so, how does the growth rate in such a regime depend on velocity ratio ?
- Does the mixing transition (Brown & Roshko 1974, Konrad 1977, Dimotakis 2000), occurring at a local Reynolds number ($Re \equiv \delta_{vis}\Delta U/\nu$, where $\delta_{vis}$ is the local layer visual thickness, $\Delta U$ is the velocity difference and $\nu$ is the kinematic viscosity) of about $10^4$ affect the mechanism underlying the growth of the shear layer, and hence also the growth rate? More generally, does small-scale structure affect large scale dynamics or growth?

- What role do the coherent structures play in the growth mechanism of the free shear layer? Does this mechanism vary with the velocity ratio across the layer?

This list is not exhaustive. The experimental and computational work reported to date has not completely resolved these questions.

The present paper is an attempt to answer the first and to some extent the second question. The third question in the list will be addressed in Part II of the paper (under preparation, preliminary results in Suryanarayanan and Narasimha, 2014). The main tool employed in this work is computation using a vortex-gas model. In this purely two-dimensional model (which has been shown to weakly converge to the smooth solutions of the 2D Euler equations, Marchioro and Pulvirenti, 1993) the vorticity is represented by a gas of point-vortices, the motion of which is entirely determined by Kelvin's theorem on the conservation of vorticity and the Biot-Savart relationship. In addition to being confined to strictly 2D turbulence, this model has the limitation of no molecule-level mixing. A justification of using such a 2D inviscid model is that the large scale coherent structures of plane free shear layers are observed to be quasi-2D in experiments (Brown & Roshko 1974, Wygnanski et al 1979), and that the spread rate is not significantly affected by viscosity even at only modestly high *Re* (see Brown & Roshko, 2012).

An extensive study of a temporally evolving free shear layer, using the vortex-gas model, was recently presented by Suryanarayanan, Narasimha and Hari Dass (2014, henceforth referred to as SNH). This is flow developing in time between two counter-flowing streams, related to the spatially evolving two-stream shear layer via Galilean transformation when the velocity difference across the layer $\Delta U = U_1 - U_2$ is small compared to the average velocity $U_m = \frac{1}{2}(U_1 + U_2)$, i.e. when $\lambda \equiv (U_1 - U_2)/(U_1 + U_2) \to 0$. The study showed that accurate simulations with sufficient number of vortices and adequate ensemble averaging can give surprisingly realistic descriptions of the large scale behavior of the flow. A regime of universal spreading rate was observed provided there is sufficient separation in scale between the size of the periodic computational domain and the length scales associated with the initial conditions. This universal spreading rate (on Galilean transformation) was noted to be within the scatter of reported self-preservation spreading rates in experiments on spatially developing mixing layers with $\lambda \lesssim 0.5$. This result, coupled with the quantitative agreement between evolution of momentum thickness in sinusoidally forced free-shear-layer experiments (Oster & Wygnanski, 1982) and appropriately initialized temporal vortex-gas simulations, suggested that the dynamics of the temporal (strictly 2D) vortex-gas was not irrelevant to describe Biot-Savart momentum dispersal in plane (2D in the mean, with 3D fluctuations) Navier-Stokes (NS) mixing layers. However, the temporal results and the conclusions strictly apply only at the shear-less ($\lambda \to 0$) limit, and there is no reason to expect them to hold in the single-stream ($\lambda \to 1$) limit. The purpose of the present study is to extend the above work on the temporal free shear layer to spatially developing free shear layers, in order to study the effect of spatial feedback or the upstream effect of downstream conditions, especially towards the single-stream limit.

It is necessary to note here that there have been several objections against using a 2D model, the most recent of which are by McMullan, Gao & Coats (2010, 2015) and D'Ovidio & Coats (2013). Based on a combination of high Reynolds number experiments and 2D and 3D LES simulations, they conclude that 2D models are inadequate to describe even the large scale evolution of post-mixing transition free shear layers. (Misleadingly, the above mentioned papers do not distinguish mixing-transition from transition to turbulence in much of their discussion.) We shall address, in this paper, the objections raised in McMullan et al (2010) regarding the capability of 2D models to predict spread rate. The issues raised by D'Ovidio & Coats (2013) and McMullan et al (2015) will be addressed in detail in Part II.

The organization of the paper is as follows. Section II sets up the problem and describes briefly the model used and the numerics. Section III considers the effect of upstream and downstream conditions on the evolution, particularly for the special and interesting case of the single-stream shear layer. Section IV discusses the effects of a wall on the high-speed side that is present in most experiments. Section V considers 'equilibrium' growth rates at different velocity ratios and compares them with experiments. We conclude with Section VI. Part II, based on much of the data of the present simulations, explores the coherent structure dynamics at different velocity ratios, and the relation between vorticity and passive scalar concentration fields, both relevant to the third question stated above.

## II. PRESENT COMPUTATIONAL SETUP

### A. Formulation and numerics

The present setup, shown in Figure 1, is an extension of the methods developed by Basu et al (1992, 1995) with one crucial improvement that we shall shortly describe. We consider a computational domain of $L \times \pm\infty$ in the $xy$ plane, containing $N_0$ point vortices of equal strength $\gamma = -L\Delta U/N_0$ at the initial instant (the initial inter-vortex spacing in $x$ being $l = L/N_0$). The numerical method used in the present work is similar to that in the temporal case (SNH), namely double precision calculations with fourth order Runge-Kutta for time integration with a time-step $\Delta t = 0.1 l/U_m$ $(= 0.1\,(2\lambda)l/\Delta U)$. Note that the Hamiltonian was conserved to within $10^{-5}$ of its initial value in temporal simulations (SNH) with $\Delta t = 0.1 l/\Delta U$. Increasing or decreasing the time-step by a factor of two is observed to change evolution of thickness by less than 2% (which is within the statistical uncertainty for the averaging time of 500 $L/U_m$ adopted here in most cases).

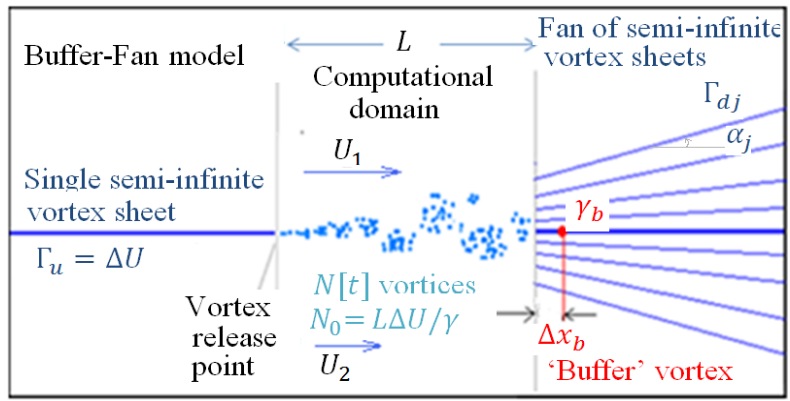

**Figure 1.** The present buffer-fan model for spatially evolving vortex-gas shear layer used throughout this work, unless specified otherwise.

In addition to the induced velocity due to the other vortices, we also include a uniform $x$-velocity of $U_m$ to ensure the $x$-velocity boundary conditions at $y = \pm\infty$ and the contributions due to the other elements present outside the domain of the present spatial setup (Fig.1), namely a semi-infinite vortex-sheet upstream, a fan of semi-infinite vortex sheets downstream (the infinite contributions cancel, see Basu et al, 1992) and a buffer vortex (whose role will be described in section IIB). Vortices enter and leave the computational domain in the spatial case. Constant strength ($\gamma$) vortices are released at $x = 0$, $y = y_i$ (drawn from a specified random distribution with amplitude $a$ usually specified in terms of $l$, or from a sinusoidal function of $t$). One such vortex is released once every $t_r$, such that ½ $(U_1 + U_2)$ $t_r \Delta U = -\gamma$. After every time-step of integration, vortices with $x > L$ are 'removed' from the computational domain. Therefore the number of vortices in the domain can fluctuate with time, $N = N(t)$. Hence the total circulation of vortices in the domain is not conserved at each instant, unlike in the temporal case - this is an important difference.

The strength per unit length of the upstream vortex-sheet is given by $\Gamma_u = \Delta U$. The strengths $\Gamma_{dj}$ and angle $\alpha_j$ made with the *x*-axis of the downstream vortex sheet $j$ in the 'fan' are chosen to approximate a symmetric, linearly spreading Gaussian vorticity profile, with a rate of spread computed from the upstream solution between $x/L = 0.2$ and $0.5$ (the reasons for this choice will be shortly clearer). We use 13 vortex sheets in the simulations reported in this work (though it is found that the solution for $x < 0.7 L$ is not significantly affected even if a single downstream vortex sheet is used for any value of $\lambda$, as will be shown in Sec IIIB). It is also ensured that $\Sigma \Gamma_{dj} \cos \alpha_j = \Delta U$. It is important to note that this model is valid only when the downstream vortex-sheets are symmetric about the *x*-axis, so as to cancel the infinite contributions to velocity from the upstream semi-infinite vortex-sheet. Therefore, this symmetry is enforced in the present simulations. The simulations reported here are started with a single downstream vortex sheet, and the fan is spread at $tU_m/L = 20$, based on the upstream solution averaged over the interval $tU_m/L = 10$ to $20$. Beyond this point, the angle of the downstream vortex sheets is dynamically varied using the cumulatively averaged upstream solution. The *x*-velocity is computed on a $0.1N \times 201$ grid and the mean velocity $\overline{U}(x, y)$ is obtained via a time-average struck over $tU_m/L = 20$ to $500$ in most of the simulations presented (except for simulations with $N_0 = 2000$, where the averages are struck over $tU_m/L = 20$ to $100$). The vorticity thickness ($\delta_\omega(x) \equiv 1/ \max_y (\partial \overline{U}/\partial y)$) is computed from such a velocity profile. The visual thickness $\delta_{vis}$ is about twice the vorticity thickness (Brown & Roshko 1974, D'Ovidio & Coats 2013) and reflects the location of the edge of the layer.

Though the above characteristics are similar in spirit to those adopted in Basu et al. (1995), the present setup differs in two ways, namely the introduction of a 'buffer vortex' that ensures instantaneous conservation of the global circulation (though Basu et al. 1995 had a 'buffer region', it did not enforce conservation of circulation), and the lack of a doublet sheet on the splitter plate. Such a sheet is necessary to ensure the physically realistic zero normal velocity on the plate. However, preliminary simulations with discrete doublet sheets revealed that they do not make any significant difference to the self-preservation spread rate (see Fig.A2 in Appendix). We also considered releasing vortices at a rate based on the instantaneous induced velocity at the tip of the plate, but the effect of such an implementation was also found to be negligible on the evolution of the layer beyond $x = 100 \, l$ (Fig. A3 in Appendix). Hence the more complex conditions at the splitter plate are not adopted in the present work as they do not justify the higher computational cost or the complexity arising due to introduction of additional parameters. The need for and the effect of the buffer vortex are described in detail in the following subsection.

## B. The 'buffer vortex'

The introduction of what we shall call 'the buffer vortex' demands careful analysis. The need for it arises due to the strong fluctuations in the number of vortices $N(t)$ in the domain. This is seen in Figure 2A which shows the RMS fluctuation ($\sigma_N$) in the number of vortices as a fraction of $N_0$ for different velocity ratios. It can be seen that $\sigma_N/N_0$ increases with $\lambda$ (and is about 12% for $\lambda = 1$) but is independent of $N$ for a given $\lambda$ (Fig.2B). To understand the origin of these fluctuations, we examine the time evolution of the number of vortices in the domain for $\lambda = 1$, shown in Fig. 2B with snapshots of vortex locations at the corresponding times shown in Fig.2C. It can be seen that there is a net accumulation of vortices during the initial transient, beyond which the number of vortices fluctuates with time around a stationary mean $<N>$, which is over 10% higher than the initial number of vortices, $N_0$. However $<N>/N_0$ is observed to be independent of $N_0$.

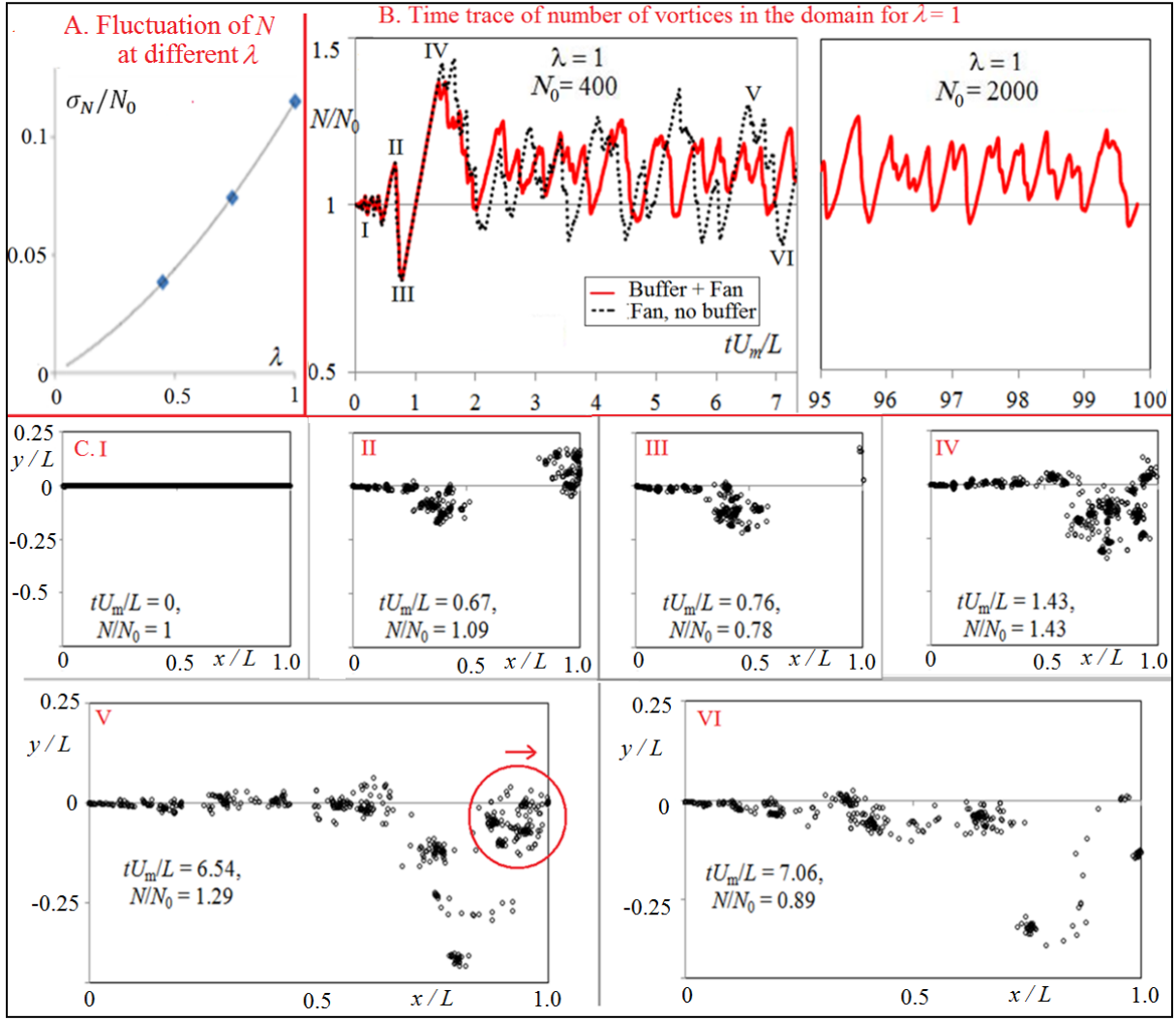

**Figure 2 (A)** RMS fluctuation of the number of vortices in the domain as a function of velocity ratio (for $N_0 = 400$, no buffer). **(B)** Time trace of the number of vortices for $\lambda = 1$. The qualitative picture remains the same on the introduction of the buffer-vortex. **C.** Snapshots of vortex-locations at times I to VI as indicated in B. For initial condition (I) in the present formulation, there is an initial transient (II – IV) that leads to a non-zero excess beyond IV due to accumulation of vortices towards the end of the domain. The fluctuations persist in the steady state as shown in V and VI. The local maxima (such as V) correspond to times just before a structure leaves the domain (see V in lower panel); and the local minima (e.g. VI) occurs immediately after such a structure has left the domain.

To understand the fluctuation about $<N>$, it is important to note that even though vortices are released at a constant rate at the edge of the splitter plate, they quickly form clusters (Fig 2B), with vortex-rich 'coherent structures' and vortex-depleted 'braid regions' between such structures. Therefore when a cluster leaves the domain, it leads to a sudden depletion of vortices in the domain, and vice versa. For example, the circled cluster exiting the domain at B leaves it with a substantial (40% of $N_0$) depletion of vortices at VI. Further, we have $\delta_\omega \sim 0.2\ x$ (for $\lambda = 1$) and the average spacing between structures is about $3\delta_\omega$. Therefore the structure passage frequency, the ratio of the convection velocity to the structure spacing, is 0.6 $L/U_m$ at $x \sim L$. The observed frequency of the fluctuation of the number of vortices in the computational domain (there are about 10 peaks in the time series of $N$ over $tU_m/L$ of 95-100 in Fig.2B, providing a fluctuation frequency of about 0.5 $L/U_m$) roughly corresponds to this passage frequency. This analysis is also supported by data at other $\lambda$ (not shown here). All this strongly suggests that the fluctuation in the number of vortices within the domain is due to the exit of vortices constituting large coherent structures from the domain.

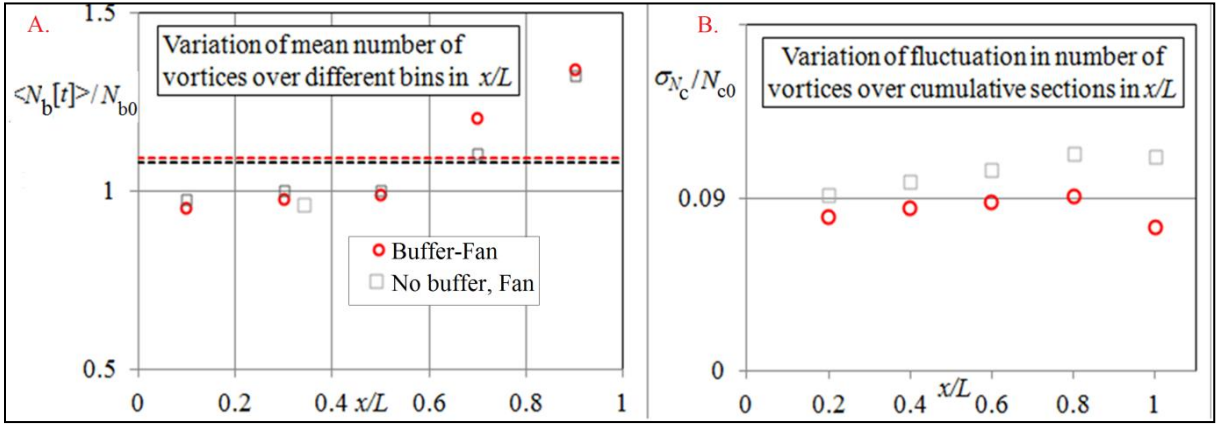

**Figure 3.** (**A**) The distribution of average number of vortices across bins of width 0.2 $L$ for $N = 2000$ (circles indicate buffer-fan results, squares simulations without buffer and the dashed lines the respective means over the entire domain). Note that for $x/L \lesssim 0.6$ there is hardly any excess over the initial value ($N_b$ refers to the number in each bin, with initial value $N_{b0} = 0.2\, N_0$ ). (**B**) Standard deviation of fluctuation of number of vortices in the region between 0 and $x$ ($\sigma_{Nc}$) as a function of $x$, normalized by the initial number of vortices in the same region ($N_{c0} = x / L\, N_0$).

While the excess over the mean number of vortices is concentrated over the last 30% of the domain (Fig.3A), the fluctuations around the mean are approximately self-similar across the domain (Fig. 3B), i.e. the fraction of the standard deviation to the mean number of vortices in the part of the domain $0<x<x_p$ hardly depends on $x_p$. Most significantly, the fluctuations are present even after the layer reaches a statistically steady state. This is an important observation because it establishes that having constant strength vortex-sheets as downstream boundary conditions leads to an instantaneous fluctuation of the total circulation of the system in time (at steady state). Therefore if the simulation is based (as is often done e.g. Basu et al 1992) on computations within the domain of length $L$ in Fig.1, without taking account of the fluctuations in the region to $+\infty$ downstream of the domain, there is in effect a violation of Kelvin's theorem. This demands an appropriate model to ensure the conservation of circulation.

One such model involves the introduction of a buffer vortex. To see how, we recall that the number of vortices in the domain dips from above the mean when a structure is just about to leave the domain, to below it when the structure has just exited (Fig.2V,VI). In reality the structure with its associated excess vorticity would then appear just downstream of the domain. Similarly the maximum surplus of vortices in the domain occurs when a vortex-depleted braid region is convected out of the domain (just preceding the exit of a structure from the domain). One would then expect a corresponding deficit immediately downstream. This cycle of excess and deficit of vorticity downstream during the deficit and excess of vortices in the domain will, because of their proximity to the downstream edge of the domain, produce associated changes in the induced velocity field upstream within the domain that must be taken into account, beyond that taken care of by the fan of vortex sheets (Fig.1). This has to be compensated for by one or more buffer regions downstream of the computational domain, as we shall shortly show that the difference with and without a correction for this fluctuation in *N* can be substantial.

The simplest procedure is to introduce a fluctuation in circulation opposite in sign to that in the domain in order to instantaneously preserve the total (global) circulation (in the infinite domain at steady state), in accordance with Kelvin's theorem. Also, since the fluctuation of vorticity in the domain should arise from the fluctuation of the vorticity out-flux downstream at *x* = *L* (as the vorticity flux from upstream is a constant), it is reasonable to introduce a fluctuation of opposite sign in the immediate downstream neighborhood of the domain. Thus, a first-order approximation to vorticity conservation would be to introduce this via a single stationary 'buffer vortex', say at $(L + \Delta x_b, 0)$, with a circulation fluctuating with time as $\gamma_b[t] = -(N[t] - N_0)\gamma$. This buffer-vortex has a non-zero (negative) mean circulation to compensate for the average excess of vortices in the domain. Nevertheless, the change in $\delta_\omega$, due to the direct contribution of the mean strength $\overline{\gamma}_b$ of the buffer vortex to the induced velocity field, is only around 2 % at *x/L* = 0.7 (for $\Delta x_b = 0, \lambda = 1$), and diminishes further upstream to less than a percent for *x/L* < 0.55. Further, it can be seen from Fig.4A that 'switching off' the mean circulation (by taking $\gamma_b = -\gamma(N - \overline{N})$) results in hardly any difference in development of the layer compared to the case with the non-zero mean. However, both cases contrast with the one without the buffer. This suggests that only the fluctuating part of $\gamma_b$ contributes significantly to altering the evolution of the layer.

The effect of the buffer vortex on the layer development increases with increase in $\lambda$. Thus, for the single-stream case shown in Fig.4, the spread rate without the buffer-vortex can be upto 30% higher over the last 70% of the domain. Thus a spread rate, determined by even a conservative fit over the first half of the domain, say over $0.1 \leq x/L \leq 0.5$, is about 0.23 for the case without a buffer vortex, but significantly lower at 0.185 for the cases where a buffer vortex has been introduced at $\Delta x_b \lesssim L$. (However, it will be shown in Sec.IIIC that the self-preservation region, which for the single-stream case without buffer vortex can end as early as 0.1 *L*, has a universal spread rate independent of the downstream boundary.) Further, it can also been seen from Fig.4B that the development of the layer hardly depends on the precise location of the buffer vortex as long as $\Delta x_b \lesssim L$.

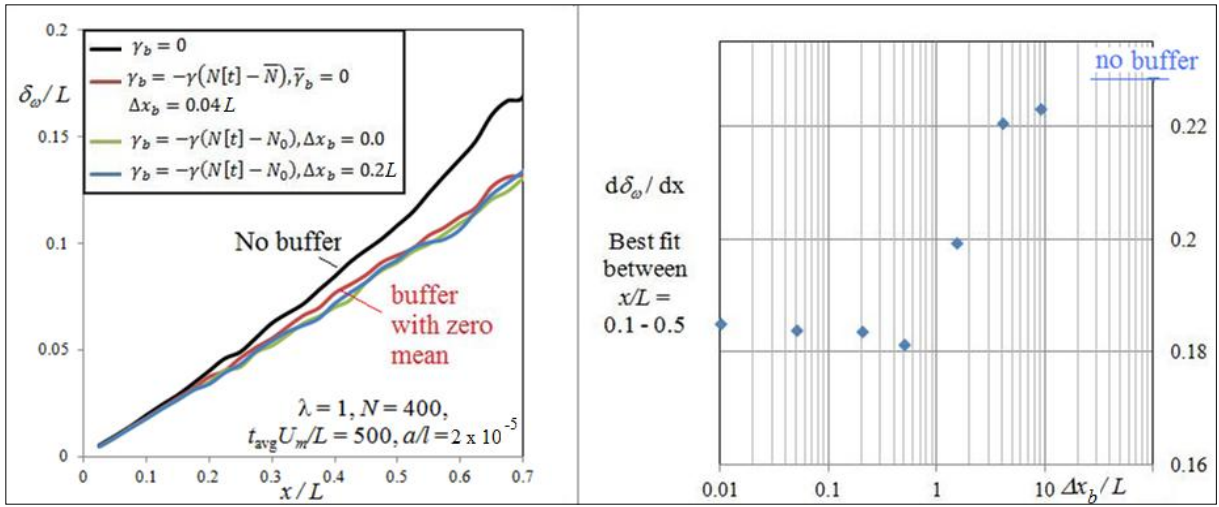

**Figure 4.** **(A)** Evolution of a single-stream vortex-gas shear layer with buffer vortex placed at different distances downstream, compared to evolution without buffer vortex. **(B)** Spread rate determined via best fit for data from *x/L* between 0.1 and 0.5.

Figure 5 compares the evolution of thickness of single-stream simulations with different domain lengths (in terms of $l$ or equivalently any other length scale associated with the initial condition scaling with $l$), and with different downstream boundaries. The solution over the region considered (i.e. $x/l <200$) for the simulation with $L/l = 2000$ is largely invariant to the downstream boundary condition as $x/L < 0.1$ (the difference between buffer-fan and no buffer with single downstream sheet is less than 2% for $x/L < 0.1$). It can be observed that the buffer-fan model with $L/l = 400$ agrees with solution of the simulation with the extended domain in the same spatial region, whereas the no-buffer case does not. This observation not only highlights the importance of the buffer vortex but also provides a validation of the buffer-fan model in simulating the upstream effects on flow development that is observed in a much larger domain, but without the associated computational cost.

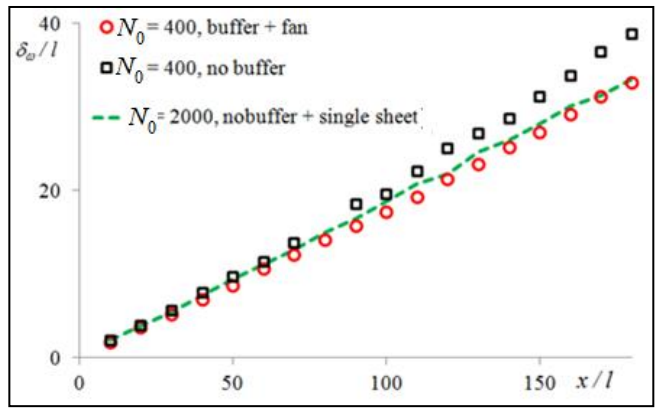

**Figure 5.** The early development of layer thickness from a simulation with and without buffer-vortex compared to simulation with larger domain (in terms of $l$).

In summary, then, the tendency of the shear layer to form vortex-dense coherent structures and vortex-sparse braid regions creates a fluctuation in the rate of vortices leaving the domain which, along with the constant vortex-release rate and static downstream boundaries, leads to fluctuation of the total circulation, leading to a violation of Kelvin's theorem. It is shown that this effect, which is most prominent in the single-stream limit, can be adequately compensated for by the introduction of an opposite signed buffer vortex downstream of the computational domain. The detailed analysis presented here, most notably comparison with simulations with a domain that is five times as long, show that a single buffer vortex with a fluctuating circulation is indeed a valid compensation for the effect of the downstream fluctuations, and is robust in the sense that the results are not very sensitive to either the mean value of circulation or the precise downstream location ($L < x_b \lesssim 2L$) of the buffer-vortex.

## III. EFFECT OF UPSTREAM AND DOWNSTREAM CONDITIONS ON THE GROWTH OF THE LAYER

In the temporal vortex-gas simulations (SNH), three regimes could be identified in the evolution of the layer: a Regime I affected by initial conditions, a Regime III influenced by the size of the finite domain and an intermediate asymptotic Regime II of universal constant spread rate unaffected by either initial conditions or domain size. This Regime II was argued to be an analog of the self-preservation or the fluid-dynamical 'equilibrium' regime (Townsend, 1956). Present simulations suggest the existence of three similar regimes in the spatial case, with the region affected by the downstream boundary being the counterpart of Regime III. Therefore, for different velocity ratios, we first examine the effect of initial conditions (both random and periodic), followed by a similar analysis of the effect of downstream boundary conditions. We then examine whether the spread rate is universal (for a given velocity ratio) in the intermediate self-preservation state (Regime II).

## A. Effect of initial conditions

We first simulate the extreme case of the single-stream shear layer for two different uniform random initial conditions with amplitudes $a$ varying by a factor of $10^5$ and with $N = 400$ and $1000$ respectively. It can be observed in Fig.6 that the two cases both attain linear growth beyond $x/l \sim 40$. Though virtual origins are different, the spread rates estimated by a best fit from $x/l = 40$ to $x/L = 0.6$ (the reason for this choice will become clearer in the next subsection) are 0.185 and 0.188 for the two cases. This is a difference of just about 1.5 percent which is within the statistical uncertainty of 2% noted in Sec.IIA. This is consistent with the observations made in the temporal layer (SNH) that the effect of initial conditions is eventually forgotten and that the spreading rate in Regime II is universal and independent of initial conditions.

The evolution of spatial vortex-gas shear layers with sinusoidal initial conditions is shown in Figure 7. In these cases, the $y$-location of the vortices released at $x = 0$ is the sum of a sinusoidal function of time and a random component. Thus, $y_i = a_w \sin[2\pi f t] + a_n Y_i$, where $a_w$ and $a_n$ are the amplitudes of the sinusoidal and random signals, $f$ is the frequency of the sinusoidal signal and $Y_i$ is a random number chosen between -1 and +1 with uniform probability. The noise component can be seen as a proxy for the free-stream turbulence level in wind tunnel experiments, something that cannot be reproduced in the vortex-gas model in any other simple way.

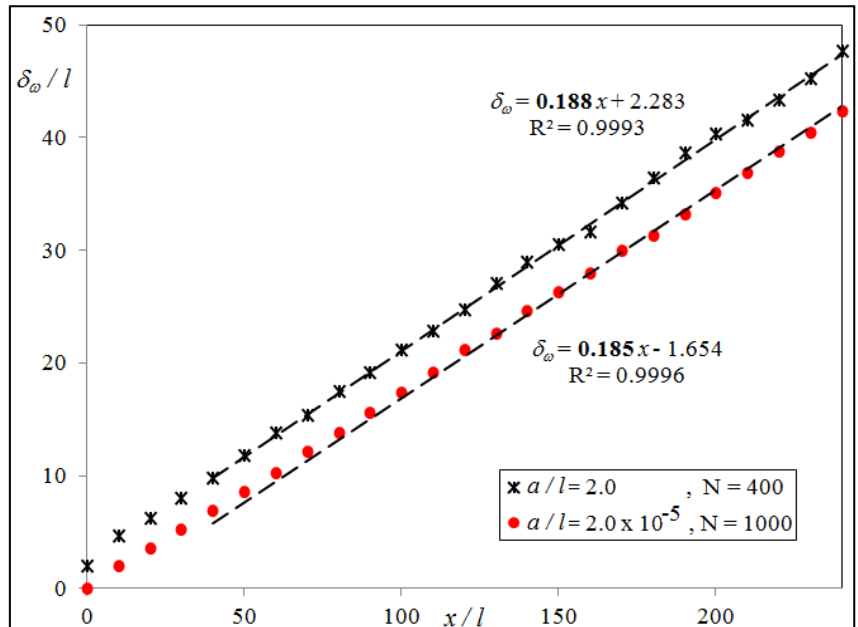

**Figure 6**. Evolution with random initial conditions with different amplitudes for $\lambda = 1$

Consistent with experiments (e.g. Oster & Wygnanski, 1982) as well as with temporal vortex-gas simulations (SNH), it can be observed from Fig.7 that the evolution is strongly affected by this periodic forcing and exhibits a strong enhancement followed by a suppression and eventual relaxation to Regime II. The spread rates in Regime II (not reached at all within the computational domain in the simulation with $\lambda = 0.11$) is observed to be within the statistical uncertainty of $\pm 2\%$ of the spread rate observed with random initial conditions for the respective values of $\lambda$, as seen in Fig.7A2 and A3. Furthermore, the simulations with lower value of $\lambda$ exhibit a longer 'memory' in terms of the wavelength of the forcing ($\Lambda = U_m/f$). It can be observed from Fig.7B that, on scaling $x$ with $\lambda\Lambda$, there is agreement in an initial stage of evolution, lasting upto the beginning of suppression for the three illustrative cases of $\lambda = 0.11$, 0.33 and 1.0. This is consistent with expectation from a Galilean transformation. However, for the same value of the noise to signal ratio, the region of suppression is reduced with increase in the value of $\lambda$, indicating that spatial feedback tends to disrupt suppression. Figure 7C shows that for $\lambda = 0.33$, there is close agreement in the evolution of thickness between the present simulations and the experiments of Oster & Wygnanski (1982) with best agreement at $a_n/a_w = 1.5$. This suggests a role of free stream turbulence or other real-world perturbations that may have been present in the experiments, play a role in the flow evolution (see SNH for more details).

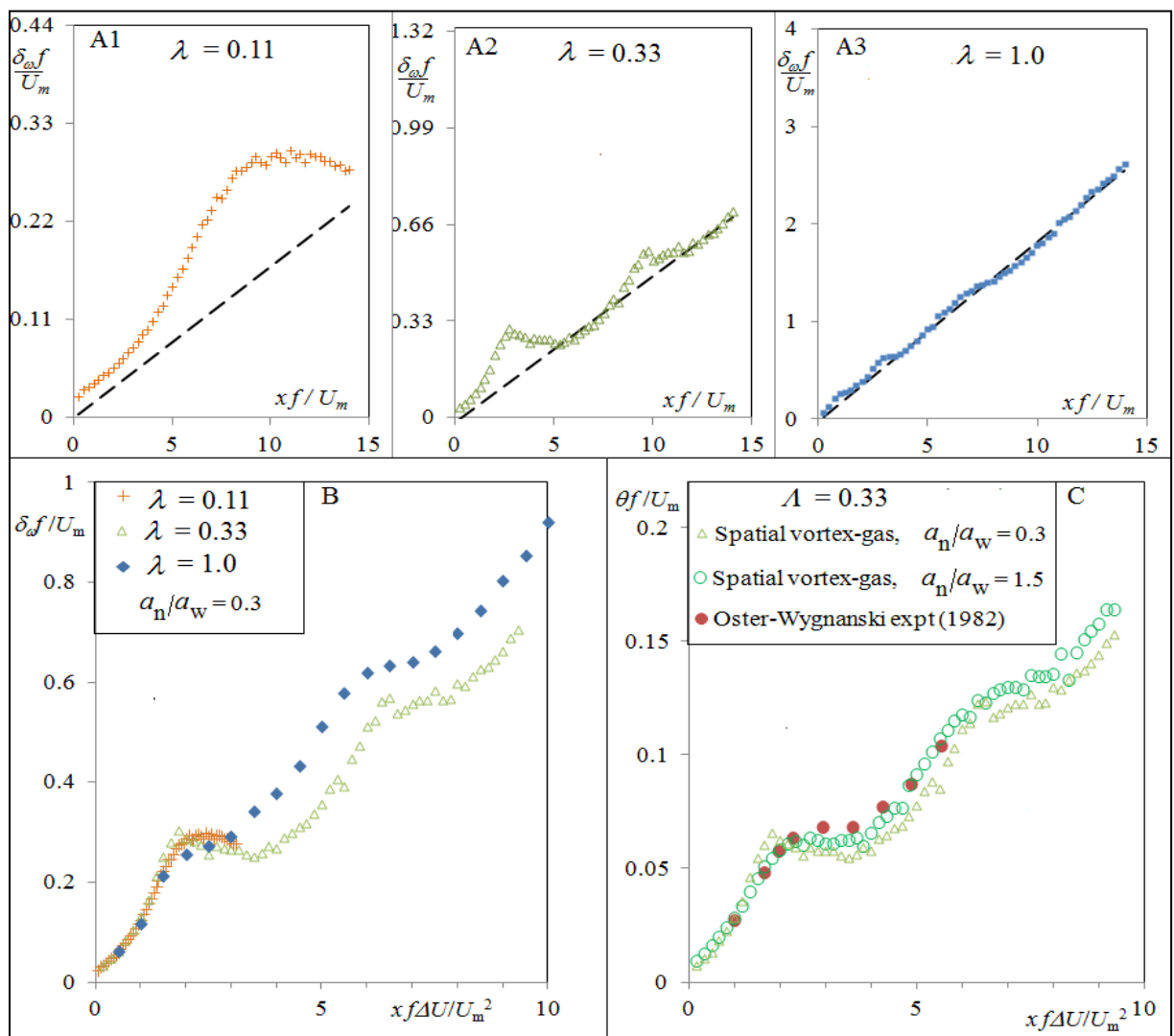

**Figure 7.** Evolution of spatial vortex-gas shear layers with sinusoidal initial vortex displacements. **A(1-3).** Evolution for different velocity ratios. **B.** Same data on scaling with wavelength of the perturbation. **C.** Effect of noise to signal ratio and comparison with experiments.

## B. Effect of the downstream conditions

We next examine, in Fig.8, the effect of downstream conditions on the evolution of the layer thickness at different velocity ratios for uniform random initial conditions. The results from $\lambda = 0.33$ with two extreme downstream boundary conditions, first with a fan of vortex sheets and a buffer vortex, and second with a single downstream vortex sheet and without a buffer vortex, are shown in Fig. 8A. It can be seen that the solution (at least in terms of thickness) is largely independent of the boundary conditions for $x \lesssim 0.8L$. This can be contrasted with the results for the same two boundary conditions shown for $\lambda = 1.0$ in Fig.8B, where the solutions for different downstream boundary conditions begin to diverge from as early as $x \approx 0.2L$.

Further, whether single vortex-sheet or a fan of vortex-sheets is used downstream, the spreading of the layer for only $x/L > 0.7$. However, the presence of a buffer vortex alters the spread rate for $x/L > 0.2$.

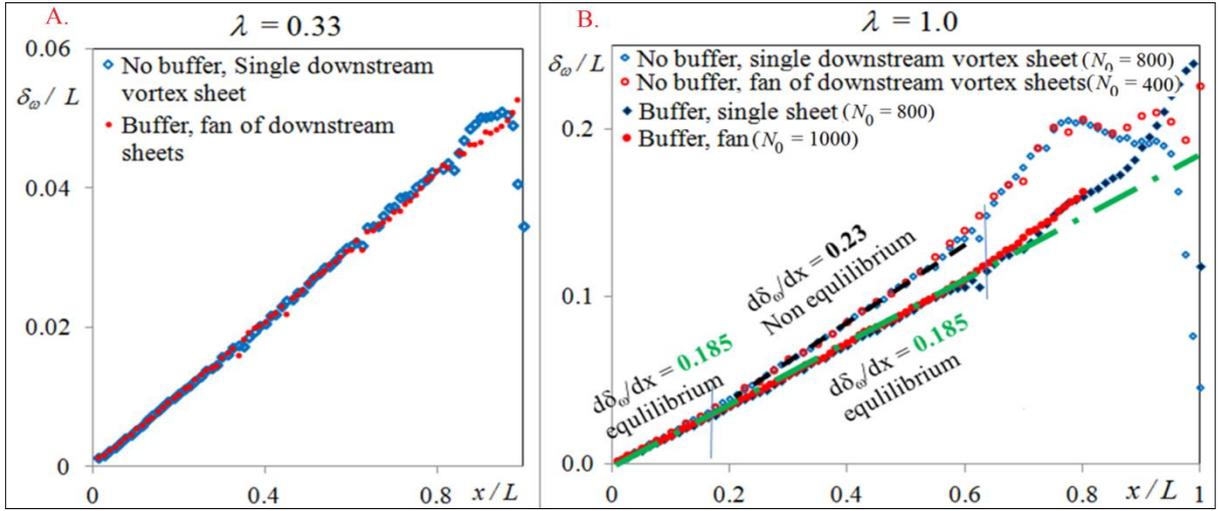

**Figure 8.** Effect of different downstream boundary conditions on evolution of the layer for A. $\lambda = 0.33$ and B. $\lambda = 1.0$.

The conclusions from Fig.8 can be summarized as follows,

- Results of simulations with the same downstream boundary condition but different $N_0$ collapse over much of the domain when scaled with $L$ – i.e. the effect of the downstream boundary scales with $L$.
- The 'zone of influence' of the downstream boundary (for the no-buffer cases) increases from under 30% of the domain for $\lambda = 0.33$ to over 80% for $\lambda = 1.0$. Thus the effect of the downstream boundary conditions becomes significantly more important with increase in $\lambda$.
- Consistent with the discussion in section IIB, the presence or otherwise of a buffer vortex (or in other words, whether or not the downstream boundary condition ensures the global conservation of circulation) makes a significant difference to the spread of the layer, over upto the last 80% of the domain in case of $\lambda = 1$.
- For certain downstream boundary conditions, such as the no-buffer cases, the spread rate estimated via a naïve fit over much of the domain (shown by the black dashed line in Fig.8) can be upto 30% larger than the buffer-fan result. Thus, a significant portion of this zone affected by downstream boundary conditions can appear to grow linearly, but with a different spread rate from that in the unaffected region. An experimental analog of this kind of downstream effect cannot be ruled out in wind tunnels or real-world scenarios, and could arise for example from interaction of the coherent structures with obstructions downstream of the measurement locations, or exit to a diffuser or the atmosphere.

## C. Universal self-preservation regime

**Table 1:** The effect of different initial and downstream boundary conditions on the self-preservation spread-rate at different $\lambda$; $x_b$ and $x_e$ denote the beginning and end of the self-preservation spread zone.

| # | Case ID | Velocity ratio ($\lambda$) | $N_0 = L/l$ | Initial condition | Downstream condition | $x_b/l$ | $x_e/L$ | $\left\|\dfrac{d\delta_\omega}{dx}\right\|_{RII}$ |
|---|---|---|---|---|---|---|---|---|
| 1 | 25_400_nbss_ur | 0.25 | 400 | Uniform random $a/l = 2 \times 10^{-5}$ | No buffer, Single-sheet | 60 | 0.75 | 0.0377 |
| 2 | 25_1000_bf_ur | 0.25 | 1000 | Uniform random $a/l = 2 \times 10^{-5}$ | Buffer – fan | 60 | 0.75 | 0.0382 |
| 3 | 33_800_nbss_ur | 0.33 | 800 | Uniform random $a/l = 0.01$ | No buffer, Single-sheet | 60 | 0.75 | 0.0534 |
| 4 | 33_800_bf_ur | 0.33 | 800 | Uniform random $a/l = 0.01$ | Buffer – fan | 60 | 0.75 | 0.0520 |
| 5 | 33_800_bf_s | 0.33 | 800 | Sinusoidal $a_w/l = 0.148$ $a_n/a_w = 1.5$ $\Lambda/l = 20$ | Buffer – fan | 240 | 0.75 | 0.0527 |
| 6 | 63_400_nbss_ur | 0.627 | 400 | Uniform random $a/l = 1.0$ | No buffer, Single-sheet | 60 | 0.7 | 0.1084 |
| 7 | 63_2000_bf_ur | 0.627 | 2000 | Uniform random $a/l = 2 \times 10^{-5}$ | Buffer – fan | 30 | 0.75 | 0.102 |
| 8 | 100_800_nbss_ur2 | 1.0 | 800 | Uniform random $a/l = 0.001$ | No buffer, Single-sheet | 10 | 0.1 | 0.187 |
| 9 | 100_2000_nbss_ur | 1.0 | 2000 | Uniform random $a/l = 2 \times 10^{-5}$ | No buffer, Single-sheet | 10 | 0.1 | 0.186 |
| 10 | 100_400_nbf_ur | 1.0 | 400 | Uniform random $a/l = 2 \times 10^{-5}$ | No buffer, fan | 10 | 0.1 | 0.188 |
| 11 | 100_800_bss_ur2 | 1.0 | 800 | Uniform random $a/l = 0.001$ | Buffer, single sheet | 10 | 0.6 | 0.182 |
| 12 | 100_1000_bf_ur | 1.0 | 1000 | Uniform random $a/l = 2 \times 10^{-5}$ | Buffer-fan | 10 | 0.6 | 0.185 |
| 13 | 100_2000_bf_ur | 1.0 | 2000 | Uniform random $a/l = 2 \times 10^{-5}$ | Buffer-fan | 10 | 0.6 | 0.188 |
| 14 | 100_400_bf_ur3 | 1.0 | 400 | Uniform random $a/l = 2.0$ | Buffer-fan | 40 | 0.6 | 0.188 |
| 15 | 100_800_bf_s | 1.0 | 800 | Sinusoidal $a_w/l = 0.296$ $a_n/a_w = 0.3$ $\Lambda/l = 40$ | Buffer-fan | 150 | 0.6 | 0.189 |

As shown in sections IIIA and B, different upstream and downstream conditions greatly alter the extent of the (unaffected or 'equilibrium') self-preservation zone. As shown in Figs. 7 and 8, the zone of influence of given initial and boundary conditions is a function of the velocity ratio. For example, as seen in Fig.8, with the buffer-fan downstream boundary condition the self-preservation extends to $x \lesssim 0.6L$, whereas without the buffer vortex, it shrinks to $x \lesssim 0.15L$ at $\lambda = 1$. However, for $\lambda = 0.33$, the self-preservation zone extends upto $x \lesssim 0.75L$, even for the no-buffer, single-sheet case. At the same time, Figs.7 and 8 strongly suggest that the spread rate itself in such a self-preservation zone seems to be unaffected by initial or boundary conditions at any given velocity ratio. We examine this issue further below.

Table 1 summarizes the results of the various simulations with different initial conditions, boundary conditions and velocity ratios. While the temporal simulations of SNH demonstrated universal self-preservation at the shear-less ($\lambda \rightarrow 0$) limit, the present study examines whether or not self-preservation is universal at higher values of $\lambda$. Taking first a look at single-stream ($\lambda$=1) cases 12 to 15, all with the buffer-fan model, a change in the initial perturbation by a factor of $10^5$ changes the growth rate by only 1.6%. This goes up 2.2% if we include the periodically forced case 15. Cases 8, 9 and 10 are within the same spread. The results of spread-rate for a wide range of initial and downstream conditions are within ±2.2% of 0.185 (the spread rate observed for case 100_1000_bf_ur, which we take as standard, as the self-preserving zone for this case extends over a decade in x and the statistics were averaged over nearly 1000 convective times). The evidence for universality at $\lambda = 1$ is thus strong. The data at $\lambda = 0.25$ and 0.33 show a difference of 1.3% and 2.7% between the maximum and minimum equilibrium spread across different cases, and thus reinforce the conclusion at $\lambda = 1$. Thus, the present results, taken together with SNH results, further support the universality of self-preservation for 2D vortex gas free shear layers (with a spread-rate that is a function of only the velocity ratio).

## IV. EFFECT OF WALL ON THE HIGH-SPEED SIDE

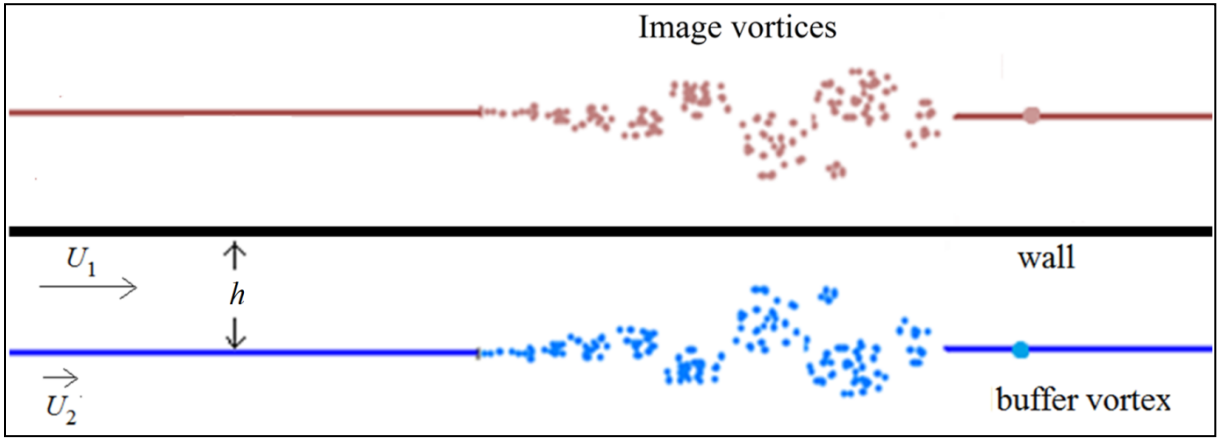

**Figure 9.** Setup for simulating spatial shear layers with a wall on the high-speed side.

The setup described in Section II considers an unbounded shear layer. However most experiments and applications have walls parallel to the splitter plate at some distance in the normal direction, often on both high and low-speed sides. Since the single-stream layer was shown in Sec. IIIB to be greatly affected by the downstream boundaries, it is important to understand its sensitivity to the presence of walls as well. In most single-stream experiments, the wall on the ambient fluid side is far away (e.g. in Hussain & Zaman 1985, the boundary on the still-fluid side is 3m away, whereas it is only 0.48 m on the flow side) and is furthermore often porous to allow for the entrainment of ambient air. On the other hand, the normal distance of wall on the flow side ($h$) from the splitter plate can be in the order of 10% of the maximum streamwise extent over which the measurements are made (e.g. 15% for Hussain & Zaman 1985, 14% for D'Ovidio & Coats 2013). Hence, in this section, we confine our attention to vortex-gas shear layers with a wall on the high-speed side, simulated with images of opposite sign. (It is worthwhile to note that Ghoniem & Ng (1987) simulate spatial vortex-gas shear layers with walls on both sides at equal distance from the splitter plate and it was observed that the wall on the low speed side dominates the interaction with the layer. Furthermore they do not consider the single stream case, which is studied in detail in this section, and hence their results are not of direct relevance here.)

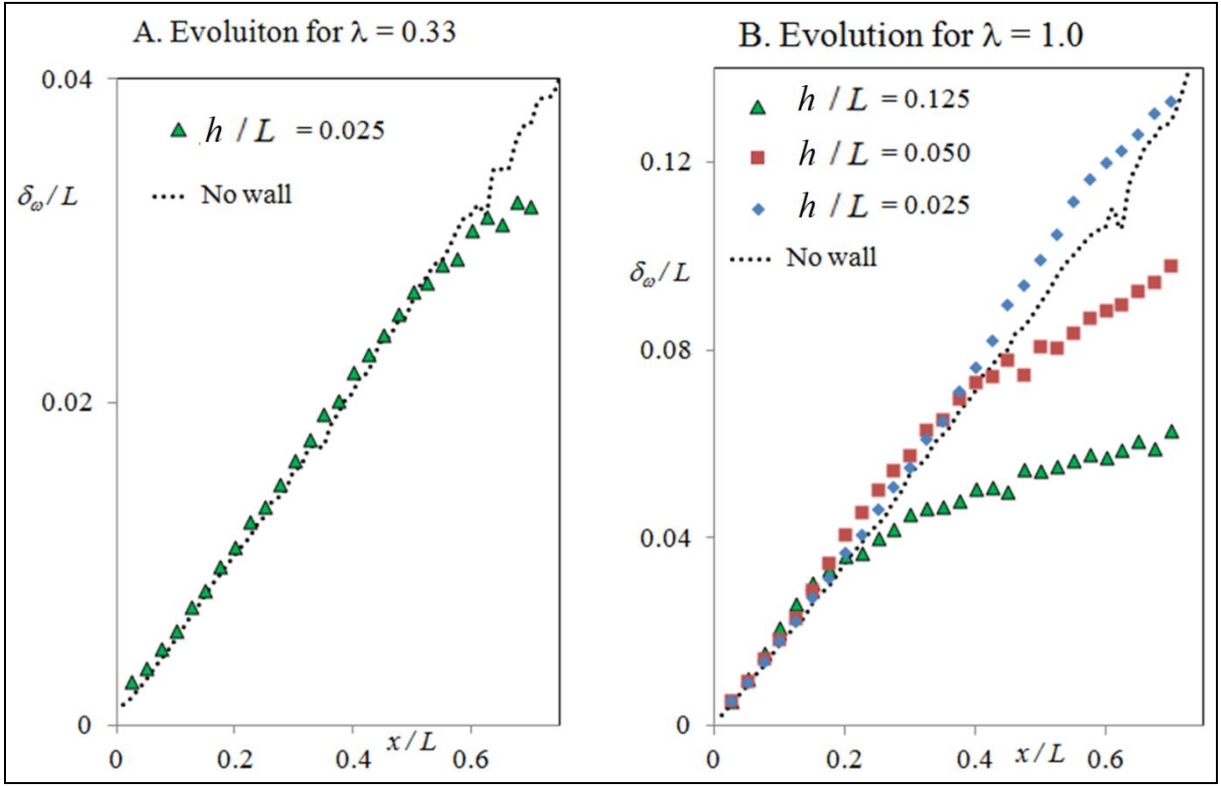

**Figure 10**. The evolution of thickness in the spatial vortex-gas free shear layers with an impermeable wall on the high-speed (/flow) side. **A.** Two stream case ($\lambda = 0.33$). **B.** Single-stream case ($\lambda = 1.0$)

The setup is illustrated in Fig. 9. The downstream boundary conditions are a single semi-infinite vortex sheet and a buffer vortex, images of both of which are also present. It is important to note that the 'wall' as simulated here ensures no penetration but permits slip. This should be a reasonable approximation to study the evolution of the free shear layer when the Reynolds numbers are sufficiently high such that the resulting boundary layers on the wall are sufficiently thin. The simulations presented in this section have $N_0 = 400$ (+ 400 images) and therefore are relatively less extensive compared to those presented in Section III. However, as we shall see below, the central picture is robust to the scatter.

Figure 10 shows the growth of the layer at two different velocity ratios $\lambda = 0.33$ and 1.0, and at different values of $h$. At $\lambda = 0.33$ (Fig. 10A) the evolution of the shear layer is largely unaffected even for $h$ as low as $0.025L$. On the other hand, at $\lambda = 1.0$ (Fig.10B), the wall exerts a significant influence at $h = 0.05L$, with the evolution of thickness strongly departing from the respective no-wall case beginning at $x/L = 0.4$. This suggests that the location of the wall with respect to the shear layer is an important factor as $\lambda \to 1$ and demands further analysis.

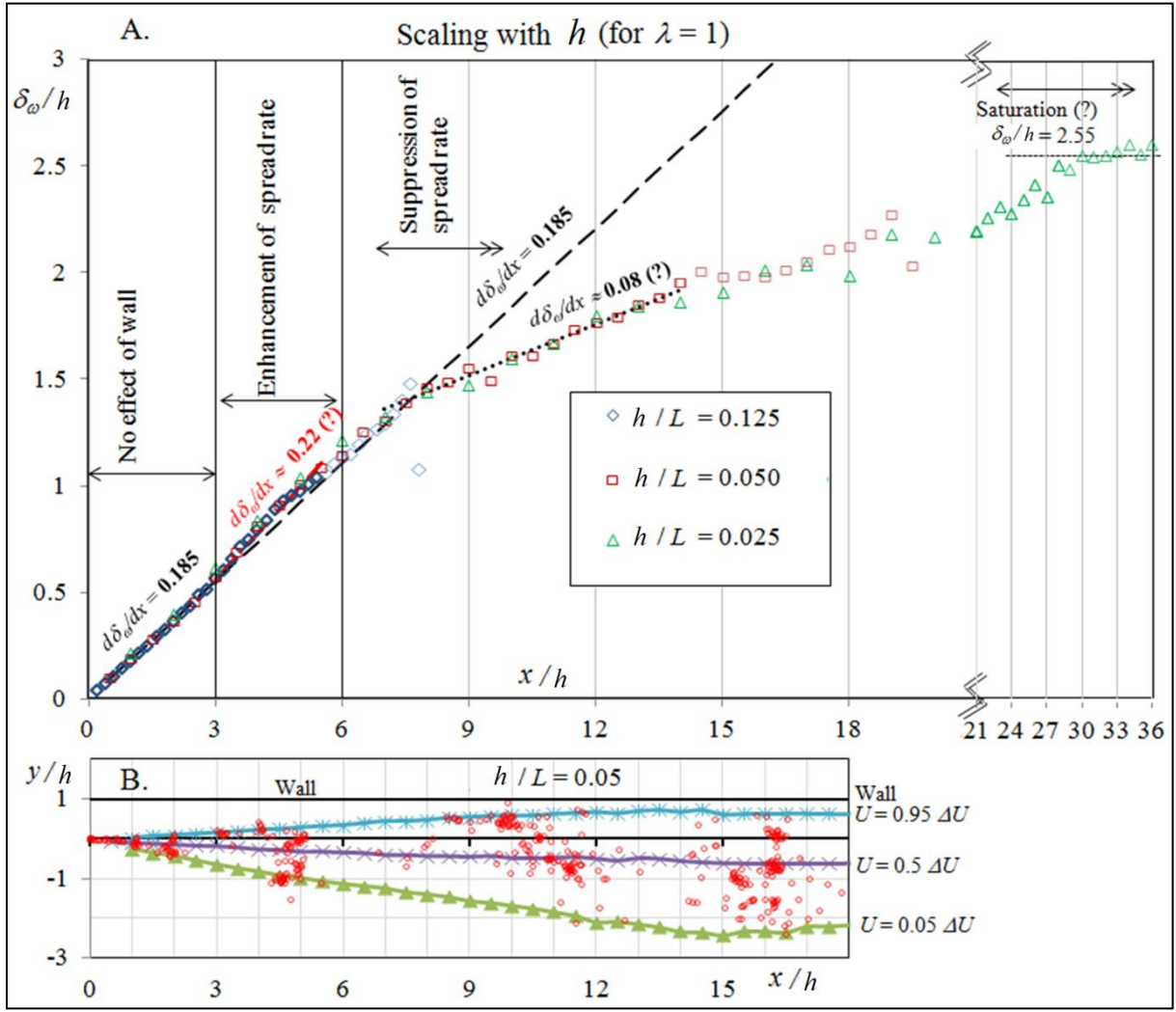

**Figure 11A.** Evolution of vorticity thickness of single stream vortex-gas shear layers, scaled with the distance to the wall on the flow side. Data from $x/L > 0.7$ are shown in faint points. (The last data point for $h/L = 0.125$ and $0.05$ is within the last 5% of the domain and hence is strongly affected by the single downstream vortex sheet). Note scaling with $h$ and the different regions in the solution (in what is a continuous function) and that the universal self-preservation is recovered only for $x/h \lesssim 3$. **B.** The evolution of $y$-locations where the velocity is 95%, 50% and 5% of the free stream velocity in relation with location of the wall, superimposed with a typical snapshot of vortex locations. (Note that $(y_{95} - y_{05}) \sim 0.75\ \delta_{vis}$).

We find from Fig.11A that the results for the three different $h/L$ roughly collapse when scaled with $h$ provided the data from initial transient and domain affected regions (latter shown in faint points in Fig.11A) are neglected. This indicates that $\delta_\omega/h$ is likely to be universal smooth function of $x/h$ in the intermediate asymptotic limit of $l \ll h \ll L$. This observation is consistent with the following argument. The effects of the wall on the layer would depend on the distance of the edge of the layer from the wall. A good measure of this distance (on neglecting the deflection of the centre of layer) is $(h - \delta_{vis}/2) = (h - \delta_\omega)$ (taking $\delta_{vis} = 2\delta_\omega$ as noted earlier). Since, in the intermediate asymptotic sense, neither $L$ nor $l$ is relevant, $\delta_\omega = F(x, (h - \delta_\omega))$, which on scaling with $h$, leads to $\delta_\omega/h = F(x/h)$.

As $\delta_\omega/h \to 0$ or equivalently $x/h \to 0$, we must recover the universal growth rate ($d\delta_\omega/dx = F' = 0.185$) and this is indeed observed in Fig.11A for $x/h \lesssim 3$. But $x > 3h$, the evolution deviates from that of the universal (no-wall) self-preservation. Though the function is continuous, it varies slowly and hence can be mistaken to be linear with different values of slope in different regions. For example, the evolution in the interval $3 < x/h < 5$, the evolution is approximately linear, and a linear fit (shown in red dashed line in Fig 11A) results in a 10% higher slope (spread rate) as compared to the universal self-preservation. The higher spread rate in the interval is seen in all three cases of $h/L$. An examination of the mean vorticity and vortex flux profiles (not shown here) suggests that, during this phase, the layer is first attracted towards the wall, leading to an increase in the vortex dispersal first on the flow side, followed by an increased dispersion on the still fluid side, which contributes to the slightly enhanced spread rate. (It has to be noted that this enhancement has not been reported in experiments (e.g. Hussain & Zaman, 1985), and will be discussed shortly.) Beyond $x \approx 6\,h$, we observe that the vorticity dispersal is greatly reduced. Again, the region between $7 \lesssim x/h \lesssim 12$, can be approximated as a linear segment, but this time with a spread rate nearly 60% less than the universal self preservation value. The rate of spread continues to reduce further with increase in $x/h$ and the present data shows that the thickness tends to saturate beyond $x \approx 30\,h$ to a value of $\delta_\omega \approx 2.55\,h$.

It can be seen from Fig.11A, for $x = 6\,h$, $\delta_\omega \sim h$ and hence $h - \delta_{\text{vis}}/2 \to 0$. Therefore the edge of the layer is already very close to the wall (as seen in Fig. 11B) and hence it is not surprising that the coherent structures and their interaction are greatly affected by the wall for $x > 6h$ leading to a suppression of layer growth.

It is important to note that this inviscid analysis provides only an outer bound of the downstream distance at which the wall begins to influence the layer evolution. In any real mixing layer at finite Reynolds number, an appropriate parameter would be the distance between the edge of the boundary layer on the wall (of thickness $\delta_{bl}$) to the edge of the shear layer, which can be estimated as $h - \delta_{\text{bl}} - \delta_{\text{vis}}/2$. As $\delta_{bl}$ and $\delta_{\text{vis}}$ are expected to increase with $x$, there could be an effect of the wall on the shear layer at an even smaller value of $x/h$ than that suggested by the present inviscid study. Furthermore, the interaction between the turbulent boundary layer on the wall and the turbulent shear layer is complex, not only involving interaction of vortical structures from the shear layer and the boundary layer which are of opposite sign, but could also be dominated by 3D and viscous effects. Therefore the detailed study of such an interaction is beyond the scope of the present 2D inviscid analysis. It cannot however be ruled out that such a complex interaction may lead to a nearly straight segments in the spread of the layer, rate of which may be different from the true self-preservation spread rate (such as in Hussain & Zaman (1985) who observe a nearly linear spread between $2 \lesssim x/h \lesssim 6.25$ but with $d\delta_\omega/dx = 0.154$). For example, it can be argued that the interaction of vortices (which are of opposite sign) from the boundary layer and the shear layer, can drive the system towards segregation into like signed vortices, an effect which could cause a suppression in layer thickness. Therefore it can be speculated that the effect arising from the interaction of the shear layer with the boundary layer may compensate for the enhancement effect observed in the present inviscid model arising from the interaction of the shear layer with its images between $3 < x/h < 5$. The combination of the two effects could thus result in a nearly linear spread, which could explain the lack of distinct observation of the enhancement in growth rate $3 < x/h < 5$ in experiments.

In summary, the effect of the walls therefore could be a major contributing factor towards variation on both sides of the quoted self-preservation spread rates across different experiments, especially in the single-stream limit, based on what is identified as the self-preserving region in each experiment.

# V. DEPENDENCE OF SELF-PRESERVATION SPREAD RATE ON VELOCITY RATIO

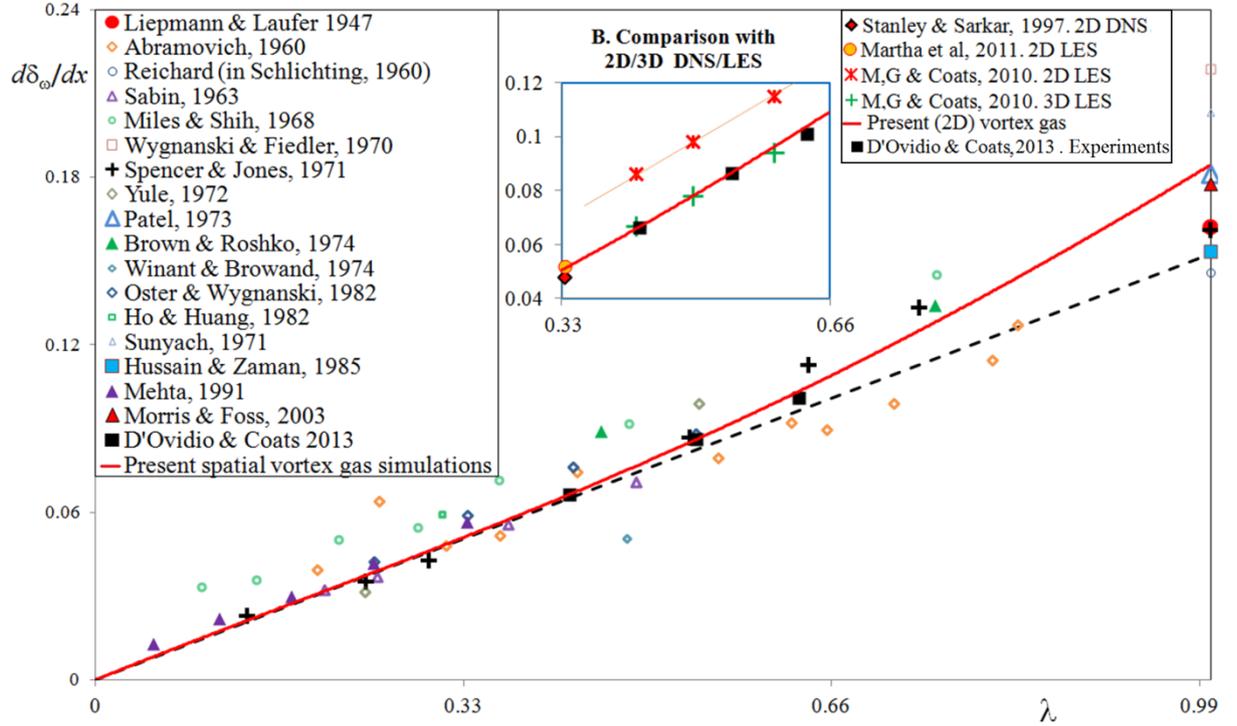

**Figure 12.** Self-preservation spread rate in the present vortex-gas simulations as a function of velocity ratio (shown as a smooth fit over data at $\lambda = 0.11, 0.25, 0.33, 0.45, 0.54, 0.63, 0.74, 0.90$ and $1.0$), compared with temporal vortex gas (SNH, dashed line), experiments and (in inset B) 2D and 3D LES and DNS studies.

Figure 12 shows the self-preservation spread rate of the present spatial vortex-gas simulations as a function of velocity ratio, and compares it with the Galilean-transformed temporal vortex-gas Regime II spread rate of SNH ($d\delta_\omega/dx = 2\lambda\, d\delta_\omega/d(t\Delta U) = 0.0153\lambda$, shown in dashed black line) as well as with nearly all available experimental data on spatially evolving mixing layers. Inset B shows the comparison with experiments (D'Ovidio & Coats, 2013), 2D and 3D Large Eddy Simulations (LES) of McMullan, Gao & Coats (2010), as well as other 2D LES/DNS studies. (The factor of 4.6, as observed for temporal vortex-gas simulations, is used to convert spread rates quoted in terms of momentum thickness, to vorticity thickness, except at $\lambda = 1$ where $\delta_\omega/\theta$ is taken as 4.8 (as inferred from the quoted spread rates of Kleis & Hussain, 1979 (in Narasimha 1990) and Hussain & Zaman, 1985). The value of 2 for $\delta_{\text{vis}}/\delta_\omega$ (as noted earlier) is used to convert ratio of the visual thickness quoted by McMullan et al (2010) into vorticity thickness.)

It is seen that for $\lambda < 0.5$ (i.e. $U_2 > U_1/3$), the spatial vortex-gas results closely agree with the SNH temporal vortex-gas simulations. This observation shows that the Galilean transformation that is exact at the shear-less limit is quite satisfactory upto a velocity difference $U_1$-$U_2$ as large as (2/3) $U_1$. Beyond $\lambda \approx 0.5$, while the experimental scatter could support either a convex or concave function, the spatial vortex-gas simulations distinctly show a mildly concave upward trend for the spread rate vs. $\lambda$. It is to be noted that this contrasts with proposals based on certain models for convective velocity (e.g. Korst et al 1955, also see Brown & Roshko 1974).

It has often been suggested (e.g. George & Davidson, 2004) that the wide scatter in the experimental data is due the existence of non-universal (i.e. initial condition dependent) self-preservation states. It could also be argued that while the present simulations demonstrate the existence of a universal self-preservation state in the 2D case, more complex, possibly 3D mechanisms in operation in real mixing layers may render them non-universal. However, it is worth noting that the present vortex-gas results are within the fairly-wide scatter of most experimental data (though, in general, there appear to be rather more experimental points above the vortex-gas result) throughout the entire range of $\lambda$. Overall, the degree of closeness between the present (2D, inviscid) vortex-gas simulations and the experiments indicate the broad relevance of the present results to a description of the large scale evolution of 3D Navier-Stokes free-shear layers at high Reynolds numbers.

## VI. DISCUSSION

In order to shed further light on the possible relevance of the present 2D model to understand 3D NS planar shear layers, we briefly analyze the observed departures of the experimental data from the present results. Figure 12 shows that the scatter is particularly large for $\lambda \to 0$ and at $\lambda = 1$. As shown in Sec. IIIA (Fig. 7A) certain initial conditions (such as those with significant long-wave periodic forcing), can have very long memories, and in terms of the physical distance (*x*), the zone of influence of a given initial condition is proportional to $\lambda^{-1}$. Therefore it is likely that scatter at small $\lambda$ is due to the memory of initial conditions, including facility-specific long-wavelength (low frequency) disturbances in the flow, not known accurately in experiments without a dominant specified forcing.

On the other hand, the present vortex-gas simulations demonstrate that the development of the single-stream layer is particularly sensitive to downstream boundary conditions. In particular, it can be seen (Fig. 8) that some downstream conditions (though numerical in this case) can affect the flow over 80% of the domain; they can also provide non-equilibrium zones that form nearly straight segments in the variation of thickness with stream-wise distance, providing a misleading estimate of self-preservation spread rates as seen in Fig.8. Furthermore, the presence of a wall on the flow side is shown to result in either higher or lower local spread rates than in the no-wall case, based on the value of *x*/*h*. (It is worth noting the experiments which are in close agreement with the present vortex-gas calculations at $\lambda = 1$, namely Patel (1973) and Morris & Foss (2003), have maximum values of *x*/*h* less than 2, a region over which universal self-preservation was observed in present simulations with wall presented in Sec IV.) Therefore, from this point of view, the large scatter at the single-stream limit is not unexpected, as many experiments may not have accounted for this extreme sensitivity to downstream conditions or walls (neither of which has been analyzed in detail earlier).

While we do not present an exhaustive analysis of individual experiments, the following provides some quick insight. It can be seen that if we only include the early experiments of Liepmann and Laufer (1947), Spencer and Jones (1971) and Brown & Roshko (1974), in all of which Reynolds numbers of over $10^5$ were reached, and all experiments post 1985, and in particular those of D'Ovidio & Coats (2013) (all of the above shown in filled symbols in Fig. 12), the scatter is significantly less and there is a closer agreement with the present results. This suggests that the large scatter across many of the early experiments is possibly due to too short or incorrectly identified self-preservation zones (as elaborated in Sections 3 and 4), rather than non-universal self-preservation. Further, the agreement of the high Reynolds number experiments with the vortex-gas results over the entire range of $\lambda$ indicates the dominance of the 2D Biot-Savart interactions in determining the large scale momentum dispersal of spatially evolving plane free shear layers at all velocity ratios.

We now address a recent assessment of the capability of 2D models to describe the large scale evolution of plane free shear layers. McMullan, Gao & Coats (2010) and D'Ovidio & Coats (2013), based on the results of high Reynolds number ($Re_{\delta\omega} \sim 3 \times 10^4$) experiments and LES, conclude that both the spread rate and growth mechanism of free shear layers post mixing transition are beyond description by 2D analysis. The inset B on Fig. 12 shows the spread rate as observed in their experiments, their 2D and 3D LES results, and the present vortex-gas simulations for $\lambda$ between 0.42 and 0.63. (Note that for this range of $\lambda$, the vortex-gas shear layer is largely insensitive to downstream conditions, see Fig 5c and Table 1.) McMullan et al (2010) reported that, while the results of their 3D LES agree with their experiments (which later appear in D'Ovidio & Coats 2013), their 2D LES predicts upto 25% higher spread rates. Based on this finding McMullan et al concluded that 2D simulations are 'wholly inadequate' for describing the large scale momentum dispersal in high Reynolds number free shear layers. However, as seen in Fig. 6, the spread rate from the present vortex-gas simulations, which are strictly 2D, agree very closely with both the experiments of D'Ovidio & Coats (2013) and the 3D LES of McMullan et al (2010). Interestingly, and paradoxically it is only the 2D LES of McMullan et al (2010) that has a significant discrepancy with the present 2D calculations and also their 3D simulations and experiments, and other 2D DNS (Stanley and Sarkar, 1997) and 2D LES studies (Martha et al, 2011) (Fig.12B).

It can be argued that the 2D LES of McMullan et al (2010) could have led to a significantly higher spread rate due to either of or a combination of the following reasons. First possibility is the choice of an initial condition involving a boundary layer on either side of the splitter plate. It is known from 2D plane jet DNS of Stanley & Sarkar (1997), supported by the preliminary simulations of a temporal vortex-gas 'wake' by Prasanth, Suryanarayanan & Narasimha (unpublished), that when vortices of opposite sign are involved, the vorticity dispersal in 2D is much higher than that in the corresponding 3D (2D in the mean) case, due to an additional contribution to momentum dispersal via vortex dipole motions away from the core flow, a mechanism that may be absent or significantly weaker when 3D fluctuations are present. The shear layer with an initial condition of two boundary layers at the trailing edge of the splitter plate is in some sense a combination of a shear layer with vortices of single sign and a wake with vortices of both signs. Our preliminary temporal simulations of such a 'shear layer plus wake' configuration (not shown here) suggest that the spread rate can be upto 20% higher over a duration (or distance in the spatial case) that is an order of magnitude longer than the duration (or distance) taken for an equivalent shear layer with a single-sign vorticity to relax to the self-preservation state. This duration, in terms of initial boundary layer thickness, is about twice the size of the domain used by McMullan et al (2010), and thus their observation of the higher spread rate within the extent of their simulation is indeed consistent with the shear layer plus wake simulations. Further analysis is being carried out to determine whether a 2D shear layer with wake will relax to the same universal spread rate as the shear layer with vorticity of one sign (though preliminary simulations appear to suggest so), and the results will be presented elsewhere. Regardless, the enhanced spread rate (whether it is an initial transient or an asymptotic effect) in the vortex-gas shear layer with vortices of both signs, appears to be due to the dipole mechanism, which as noted would be much weaker in a 3D setting. This could explain why the paradox of why 3D LES of McMullan et al (2010), even with the same shear layer plus wake initial condition as in the 2D case, is in agreement with the present (single sign initial condition) vortex gas results and experiments, their 2D LES is not. Further supporting this argument is the agreement in spread rate (see Fig.12B) with the present simulations of the 2D LES study of Martha et al (2011), which used single sign vorticity as initial condition. (However, it must be noted that the 3D LES of Martha et al (2011) show much higher spread rates than their 2D case or experiments. This difference, however, is in a direction opposite to that observed by McMullan et al (2010). As the authors themselves suggest, the higher growth rate in their 3D simulations appears to be the choice of their initial condition of 3D perturbations involving counter rotating streamwise vortices in their 3D LES, which are presumably not forgotten within the extent of their computational domain.)

Secondly, 2D LES calculations often use standard sub-grid models, which have been developed and tuned to predominantly handle 3D flows (e.g. McMullan et al 2010 use standard Smagorinsky model with the same value of Smagorinksy constant $C_s = 0.12$ for both their 2D and 3D simulations). The rigor of this approach can be questioned considering the very different small scale behavior of 2D and 3D turbulence (e.g. Davidson, 2004).

Finally, it is important to note that averaging times adopted by McMullan et al (2010) as well as in other simulations (such as Martha et al, 2011) are O(10 $L/U_m$) and so relatively short, as this can be contrasted with the present simulations where statistics are averaged over 100 to 1000 $L/U_m$ depending on the case). It can be inferred from the present simulations that the uncertainty associated with such short averaging times can be order of $\pm 10\%$, and therefore extreme caution is needed in evaluating them based on closeness of their agreement with experiments (or the present simulations).

Regardless of why predictions of the 2D LES of McMullan et al (2010) are not in agreement, the suggestion that 2D simulations are inadequate in post-mixing transition shear layers is contradicted by the remarkable agreement between the present 2D vortex-gas simulations (with single sign vorticity) with both high *Re* experiments and 3D LES, both post-mixing transition. Based on the analysis of coherent structures, D'Ovidio & Coats (2013) argue that the growth mechanism of post-mixing transition shear layers are greatly different from those of pre-mixing transition layers and likely to be 3D. This raises the question of how the spread rates of high Reynolds number experiments including the post-mixing-transition cases of D'Ovidio & Coats (2013) are accurately predicted by 2D models such as the present vortex-gas simulations. This issue is addressed in part II, where we present a detailed analysis of coherent-structure dynamics in vortex-gas simulations of the kind reported here.

## VIII. CONCLUSION

The present spatially evolving vortex-gas simulations show that the extent of the upstream influence of the downstream boundary conditions increases with $\lambda$, and varies from less than 20% of the domain length at $\lambda = 0.33$ to over 80% at $\lambda = 1$. A simple explanation for the longer upstream influence at $\lambda = 1$ is the absence of the mean streamwise advective velocity on one side of the free shear layer. However, under conditions of sufficiently long flow development for initial conditions to be forgotten, a measurement zone sufficiently far from downstream boundaries, and no interference from a top wall, the spread rate is a universal concave-upward function of velocity ratio parameter $\lambda$. Agreement of present 2D vortex-gas simulations with high Reynolds number (post-mixing transition) experiments and 3D LES (McMullan et al, 2010), over a range of velocity ratios, suggests that spreading by a 2D mechanism provides adequate prediction of the 3D flow over the whole range of $\lambda$.

# ACKNOWLEDGEMENTS

We thank Prof. Garry Brown (Princeton) and Prof. Anatol Roshko (Caltech) for many rewarding and enjoyable discussions and suggestions. We also thank Mr. Gokul Pathikonda (Univ. of Illinois at Urbana-Champaign) for help with early analysis during the initial part of this work. We acknowledge support from DRDO through the project RN/DRDO/4124 and from Intel through project RN/INTEL/ 4288.

**Appendix : Supplementary data**

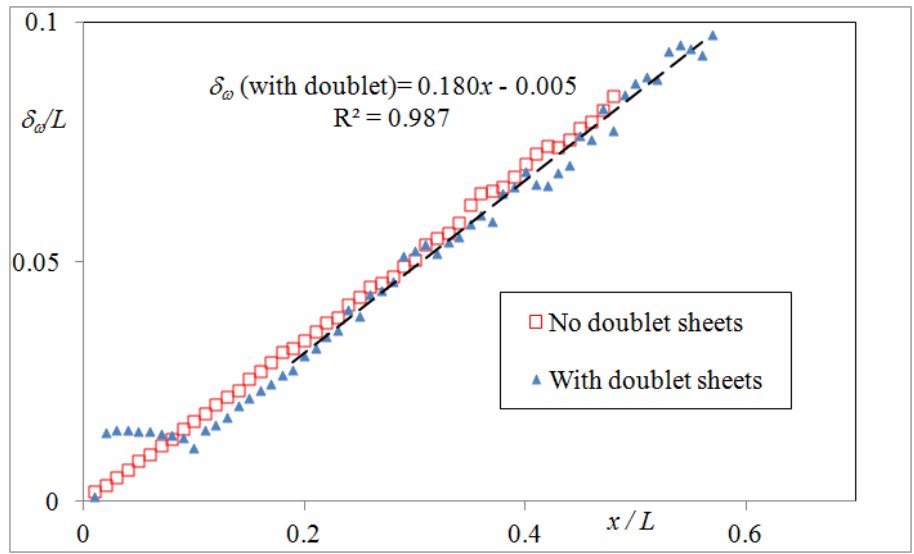

**Figure A1**. The robustness of self-preservation spread rate to presence of doublet sheets on splitter plate (at $\lambda =1$).

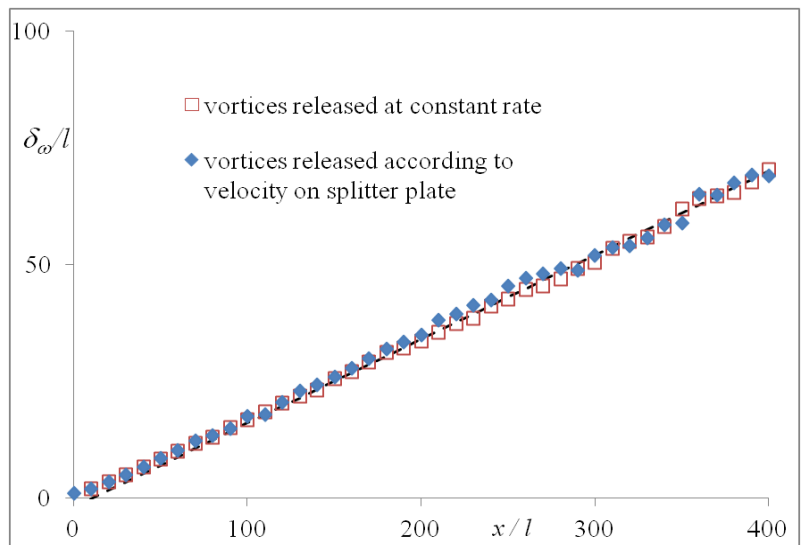

**Figure A2.** The robustness of self-preservation spread rate on constant or variable rate of vortex-release at the end of the splitter plate (at $\lambda =1$).